\input harvmac.tex
\vskip 2in
\Title{\vbox{\baselineskip12pt
\hbox to \hsize{\hfill}
\hbox to \hsize{\hfill}}}
{\vbox{\centerline{Ghost Cohomologies and Hidden Space-Time Symmetries}
\vskip 0.3in
{\vbox{\centerline{}}}}}
\centerline{Dimitri Polyakov\footnote{$^\dagger$}
{dp02@aub.edu.lb}}
\medskip
\centerline{\it Center for Advanced Mathematical Studies}
\centerline{\it and  Department of Physics }
\centerline{\it  American University of Beirut}
\centerline{\it Beirut, Lebanon}
\vskip .5in
\centerline {\bf Abstract}
We observe and study new non-linear global
space-time symmetries of the full ghost$+$matter action of
$RNS$ superstring theory.
We show that these surprising new symmetries are generated by
the special worldsheet currents ( physical vertex operators)
of RNS superstring theory, violating the equivalence
of superconformal ghost pictures.
We review the questions of BRST invariance and nontriviality
of picture-dependent vertex operators
and show their relation to hidden space-time symmetries and
hidden space-time dimensions.
In particular, we relate 
the space-time transformations, induced by picture-dependent currents,
to the symmetries observed in the 2T physics approach.
 {\bf }
\Date{January 2007}
\vfill\eject
\lref\verl{H. Verlinde, Phys.Lett. B192:95(1987)}
\lref\bars{I. Bars, Phys. Rev. D59:045019(1999)}
\lref\barss{I. Bars, C. Deliduman, D. Minic, Phys.Rev.D59:125004(1999)}
\lref\barsss{I. Bars, C. Deliduman, D. Minic, Phys.Lett.B457:275-284(1999)}
\lref\lian{B. Lian, G. Zuckerman, Phys.Lett. B254 (1991) 417}
\lref\pol{I. Klebanov, A. M. Polyakov, Mod.Phys.Lett.A6:3273-3281}
\lref\wit{E. Witten, Nucl.Phys.B373:187-213  (1992)}
\lref\grig{M. Grigorescu, math-ph/0007033, Stud. Cercetari Fiz.36:3 (1984)}
\lref\witten{E. Witten,hep-th/0312171, Commun. Math. Phys.252:189  (2004)}
\lref\wb{N. Berkovits, E. Witten, hep-th/0406051, JHEP 0408:009 (2004)}
\lref\zam{A. Zamolodchikov and Al. Zamolodchikov,
Nucl.Phys.B477, 577 (1996)}
\lref\mars{J. Marsden, A. Weinstein, Physica 7D (1983) 305-323}
\lref\arnold{V. I. Arnold,''Geometrie Differentielle des Groupes de Lie'',
Ann. Inst. Fourier, Grenoble 16, 1 (1966),319-361}
\lref\self{D. Polyakov,  Int. Jour. Mod. Phys A20: 2603-2624 (2005)}
\lref\selff{D. Polyakov, Phys. Rev. D65: 084041 (2002)} 
\lref\sellf{D. Polyakov, Int. J. Mod. Phys A20:4001-4020 (2005)}
\lref\selfian{I.I. Kogan, D. Polyakov, Int.J.Mod.PhysA18:1827(2003)}
\lref\doug{M.Douglas et.al. , hep-th/0307195}
\lref\dorn{H. Dorn, H. J. Otto, Nucl. Phys. B429,375 (1994)}
\lref\prakash{J. S. Prakash, 
H. S. Sharatchandra, J. Math. Phys.37:6530-6569 (1996)}
\lref\dress{I. R. Klebanov, I. I. Kogan, A. M.Polyakov,
Phys. Rev. Lett.71:3243-3246 (1993)}
\lref\selfdisc{ D. Polyakov, hep-th/0602209, to appear
in IJMPA}

\centerline{\bf Introduction}

In RNS superstring theory the physical states are described
by BRST-invariant and nontrivial vertex operators, corresponding to various
string excitations. Typically, these operators
are defined up to  transformations by the picture-changing operator
$\Gamma=\lbrace{Q_{brst},\xi}\rbrace$ and its inverse
$\Gamma^{-1}=-4c\partial\xi{e^{-2\phi}}$ where $\xi=e^\chi$ and 
$\phi,\chi$ is the pair of the bosonized superconformal ghosts:
$\beta=\partial\xi{e^{-\phi}}$ and $\gamma=e^{\phi-\chi}$.
Acting with $\Gamma$ or $\Gamma^{-1}$ changes the ghost 
number of the operator by 1 unit, 
therefore each perturbative string excitation (such as a photon or
a graviton) can be described by an infinite set of physically equivalent
operators differing by their ghost numbers, or the ghost pictures.
Typically, for a picture n operator one has
$$:\Gamma{V^{(n)}}:={V^{(n+1)}}+\lbrace{Q_{brst},...}\rbrace$$
and
$$ :\Gamma^{-1}V^{(n)}:=V^{(n-1)}+\lbrace{Q_{brst},...}\rbrace$$
The inverse and direct picture-changing operators satisfy the OPE identity
\eqn\grav{\eqalign{\Gamma(z)\Gamma^{-1}(w)=1+\lbrace{Q}_{brst},\Lambda(z,w)
\rbrace\cr
\Lambda(z,w)=:\Gamma^{-1}(z)(\xi(w)-\xi(z)):}}
The standard perturbative superstring vertices can thus be represented
at any integer (positive or negative) ghost picture.
Such a discrete picture changing symmetry is the consequence
of the discrete automorphism symmetry in the space of the supermoduli
 ~{\verl},~{\self}. Varying the location of picture-changing operator
(or, equivalently, varying the super Beltrami basis)
inside correlation functions changes them by the full derivative
in the supermoduli space.
 This ensures their picture invariance
after the appropriate moduli integration, if the supermoduli
space has no boundaries or global singularities.
The global singularities, however, do appear in case if
a correlation function contains vertex operators $V(z_n)$ such that the
supermoduli coordinates diverge faster 
than $(z-z_n)^{-2}$ when they approach  the insertion points
 on the worldsheet
~{\self}. If the latter is the case, the moduli 
integration  of the full derivative term  would result
in the nonzero boundary contribution and the correlation function
would be picture-dependent.
The first important physical example of how  
 picture-dependent operators emerge in superstring theory
  is the following.
Consider NSR superstring theory (critical or noncritical)
in $d$ dimensions. 
Its wordlsheet action is given by
\eqn\grav{\eqalign{S={1\over{2\pi}}\int{d^2z}\lbrace
-{1\over2}{\partial{X_m}}\bar\partial{X^m}-{1\over2}
\psi_m\bar\partial\psi^m-{1\over2}\bar\psi_m\partial\bar\psi^m
+b\bar\partial{c}+{\bar{b}}\partial{\bar{c}}+
\beta\bar\partial\gamma+\bar\beta\partial\bar\gamma\rbrace}}
where, for a time being, we have skipped the Liouville part.
This action is obviously invariant under two global
$d$-dimensional space time symmetries -
Lorenz rotations and translations. 
It does have, however, yet another surprising 
and non-trivial space-time symmetry,
which, to our knowledge, has not been discussed in the literature so far.
That is, one can straightforwardly  check that
the action (2) is invariant under the following non-linear
global transformations, mixing the matter and the ghost sectors of the
theory:

\eqn\grav{\eqalign{\delta{X^m}=\epsilon
\lbrace\partial(e^\phi\psi^m)+2e^\phi\partial\psi^m\rbrace\cr
\delta\psi^m=\epsilon\lbrace{-}e^\phi\partial^2{X^m}
-2\partial(e^\phi\partial{X^m})\rbrace\cr
\delta\gamma=\epsilon{e^{2\phi-\chi}}(\psi_m\partial^2{X^m}
-2\partial\psi_m\partial{X^m})\cr
\delta\beta=\delta{b}=\delta{c}=0}}

The accurate proof of this fact will be given in the 
section 3 of this paper.
Given the transformations (3),it is not difficult to check that
their  generator is given by

\eqn\lowen{T=\int{{dz}\over{2i\pi}}{e^\phi}(\partial^2{X_m}\psi^m
-2\partial{X_m}\partial\psi^m)}

The integrand of (4) is a primary field of dimension 1,
i.e. a physical generator.
While it is not manifestly BRST-invariant (it
doesn't commute with the supercurrent terms of $Q_{brst}$)
below it will be demonstrated how its BRST invariance can be restored by 
adding the appropriate b-c ghost
dependent  terms.
The peculiar property of this generator is that 
it is annihilated by $\Gamma^{-1}$ and
 has no analogues
at higher pictures, such as $0,-1$ and $-2$. 
There is, however, a picture $-3$ version of this
generator, with the manifest BRST invariance. This version can be 
obtained simply by replacing $e^\phi\rightarrow{e^{-3\phi}}$ in (4).
Similarly to the picture $+1$-version, the picture ${-3}$
version is annihilated by $\Gamma$, so the operator is 
picture-dependent.
We will return to the operators of the type (4),
as well as to the surprising symmetry (3), in the next sections
of the paper, and for now will consider
the next important example of picture-dependent vertex operators
 in critical RNS superstring theory,given by 
 the five-form state 
\eqn\grav{\eqalign{
V_5(k)=H(k)\oint{{dz}\over{2i\pi}}U_5(z)\equiv
H_{m_1...m_5}(k)\oint{{dz}\over{2i\pi}}
e^{-3\phi}\psi^{m_1}...\psi^{m_5}e^{ikX}(z)}}
with the five-form $H$
subject to the BRST non-triviality condition
\eqn\lowen{k_{{\lbrack}m_6}H_{m_1...m_5\rbrack}(k)\neq{0}}
or simply $dH\neq{0}$.
That is, if the condition (6) isn't satisfied,
there exists an operator
\eqn\lowen{
S_5=H_{m_1...m_5}(k)\oint{{dz}\over{2i\pi}}
\partial\xi{e^{-4\phi}}\psi^{m_1}...\psi^{m_5}(\psi^m\partial{X_m})
e^{ikX}}
which commutator with $Q_{brst}$ is $V_5$, i.e. the five-form (5)
is trivial. Indeed, if the $H$-form is closed, the $S_5$ operator  (7)
is the primary field and its commutator with $\gamma\psi\partial{X}$
term of the BRST charge gives $V_5$. If, however,
the condition (6) is satisfied, $S_5$ doesn't commute with the stress tensor
$cT$ term of $Q_{brst}$ and therefore $V_5$ cannot be represented
as a commutator of $Q$ with $S_5$ or any other operator, as we will show
by explicit computation of its correlators.
The non-triviality condition (6) has a natural physical interpretation.
It simply implies that all the nonzero scattering amplitudes
involving the $V_5$-operator (5) must be the functions of $dH$
rather than  $H$ itself; accordingly, they vanish 
for the closed $H$-forms when the $V_5$-operator becomes trivial.
This $V_5$ operator is annihilated by the picture-changing transformation
and the supermoduli approaching $U(z)$ diverge as 
${(z-w)}^{-4}$ which can be read off the OPE of $U$ with the worldsheet
 supercurrent: 
\eqn\lowen{G(w)U_5(z)\sim(z-w)^{-4}c\xi{e^{-4\phi}}\psi^{m_1}...\psi^{m_5}
e^{ikX}+...}
 As it will be shown below, the correlation functions involving the 
operators of the $V_5$-type are picture-dependent and in fact this
is one of the direct consequences of (8).
The $V_5$-operator of (5) is an example of the ghost-matter mixing operator
as its dependence on the superconformal ghost degrees of freedom
can't be removed by the picture-changing  and therefore isn't just
an artifact of the gauge. This operator exists at all the negative pictures
below $-3$, but has no version at pictures $-2$, $-1$ or zero.
At the first glance the existence of such a physical operator
seems to contradict our standard understanding of the picture-changing,
implying that all the vertices must exist at all equivalent pictures.
So the first suspicion is that the $V_5$-state is BRST-trivial,
as the relation (1) seems to imply that all the BRST-invariant operators
annihilated by picture-changing can be written as $\lbrack{Q_{brst},S}\rbrack$
for some S. 
Indeed, if $V_5$ is annihilated by $\Gamma$,
 using (1) and the invariance of $V_5$, 
we have
\eqn\lowen{0\equiv{lim}_{w\rightarrow{z}}\Gamma^{-1}(u)\Gamma(w)V_5(z)
=1+{lim}_{w\rightarrow{z}}\lbrace{Q_{brst},\Lambda(u,w)}{V_5(z)}\rbrace}
and hence $V_5$ is the BRST commutator:
\eqn\lowen{V_5(z)=-lim_{w\rightarrow{z}}\lbrace{Q_{brst}},
\Lambda(u,w)V_5(z)\rbrace}
This commutator can be written as
\eqn\grav{\eqalign{lim_{w\rightarrow{z}}\lbrace{Q_{brst}},
\Lambda(u,w)V_5(z)\rbrace=lim_{w\rightarrow{z}}\lbrace{Q_{brst}},
\Gamma^{-1}(u)(\xi(w)-\xi(u))V_5(z)\rbrace\cr
=lim_{w\rightarrow{z}}\Gamma^{-1}(u)(\Gamma(w)-\Gamma(u))V_5(z)}}
where we used $\lbrace{Q_{brst},\xi}\rbrace=\Gamma$
and $\lbrack{Q_{brst},\Gamma^{-1}}\rbrack=0$.
But since $V_5$ is annihilated by $\Gamma$ at coincident points,
\eqn\lowen{lim_{w\rightarrow{z}}\Gamma^{-1}(u)\Gamma(w)V_5(z)=0}
and the commutator (10) is given by
\eqn\lowen{V_5(z)=\lbrace{Q_{brst}},
{\xi}\Gamma^{-1}(u)V_5(z)\rbrace}
Thus  $V_5$  is indeed the BRST commutator, but with an operator
$outside$ the small Hilbert space and therefore (1) does not by
itself imply its triviality.
This point is actually quite a conceptual one.
The very possibility of  having
 BRST non-trivial and invariant operators annihilated by $\Gamma$
(or similarly, by $\Gamma^{-1}$) follows from the fact that
the bi-local $\Lambda(u,w)$-operator (which, by itself,
is $in$ the small Hilbert space)
 vanishes at coincident points.
Had this not been the case, (1) would have implied that
any operator $V$, annihilated by $\Gamma$,
would have been given by the commutator
$V(z)=-\lbrace{Q_{brst}},\Lambda(z,z)V\rbrace$,
i.e. would have been BRST exact.
But since $\Lambda(z,z)={0}$, the OPE (1)
only leads to the trivial identity of the type $V=V$,
as it is clear from (13) (note that ${\lbrace}Q_{brst},\xi\Gamma^{-1}\rbrace
=1$)
The picture-dependent 
$V_5$ operator (5) is actually BRST non-trivial and invariant,
i.e. it is a physical state.
In fact, the $V_5$-operator is not related to any point-like perturbative
string excitation but to the D-brane dynamics ~{\selfian}, and its non-perturbative
 characted is somehow encripted in its non-trivial picture dependence.
While the questions of its non-triviality have been discussed previously
(see, e.g. ~{\selfian}),
 below we shall review the standard arguments leading
to the standard concept of picture equivalence of physical states
(which indeed is true for the usual perturbative point-like string excitations
like a graviton or a photon) and show where
precisely these arguments fail for the case of $V_5$.
So let us further 
analyze the question of the nontriviality of picture-dependent 
operators.
The next
 standard argument for the picture-equivalence
stems from the  fact that one is able to freely move the picture-changing
operators inside the correlators.
Indeed, since the derivatives of $\Gamma$ are all BRST-trivial:
$\partial^n{\Gamma}=\lbrace{Q_{brst}},\partial^n\xi\rbrace;n=1,2,...$
one can write 
\eqn\lowen{\Gamma(w)=\Gamma(z)+\lbrace{Q_{brst}},
\sum_n{{(w-z)^n}\over{n!}}\partial^n\xi\rbrace.}
Next, using the identity $1=:\Gamma\Gamma^{-1}:$,
following directly from (1), one can write for any correlator
including $V_5$ and some other physical operators $U_i(z_i),i=1,...n$
\eqn\lowen{
<V_5(z)U_1(z_1)....U_n(z_n)>=<\Gamma^{-1}\Gamma(w)V_5(z)U_1...U_n>}
for any poinnt $w$. Using (14) one can move $\Gamma$ from $w$ to $z$
but, since $:\Gamma{U_5}:(z)=0$ this may seem to imply that the correlator
(15) is zero, bringing about the suspicion for the BRST triviality
of $V_5$
 However, a more careful look at the problem shows that
this argument only implies the vanishing
of some (but not all) of the correlation functions
of $V_5$ but, strictly speaking, says nothing about its BRST triviality.
That this argument does not imply the vanishing of $all$ the
correlators of $V_5$, can be shown from the following.
The non-singlular OPE of $\Gamma$ with $U_5$ 
 is given by
\eqn\lowen{\Gamma(z)U_5(w)\sim(z-w)^2{e^{-2\phi}}\psi^{{\lbrack}m_1}
...\psi^{m_4}
(i(k\psi)\psi^{m_5\rbrack}+\partial{X_{m_5\rbrack}})e^{ikX}}
(here we imply the antisymmetrization
of over the space-time indices and
ignore the $c\partial\xi$ term of $\Gamma$ which is irrelevant
for integrated vertices).
This operator product vanishes for $z=w$.
Imagine, however, that the correlation function involving $U_5$
also contains another integrated physical operator,
$\oint{{du}\over{2i\pi}}W_5(u)$ such that the OPE of 
$\Gamma$ and $W_5$ is $singular$ and the singularity order
is greater, or equal to $(z-u)^{-2}$. Suppose
such a singularity $cannot$ be removed by the picture-changing
so it is not the artifact of a picture choice
(e.g. as this would be the case for a photon at picture $+1$).
 Then integral over $u$
particularly includes the vicinity of $z$, the location of $U_5$.
If $lim_{u\rightarrow{w}}\Gamma(z)W_5(u)\sim(z-w)^{-n};n\geq{2}$,
moving the location of $\Gamma$ from $z$ to $w$ can no longer annihilate
the correlation function and hence corresponding amplitude will be nonzero.
Thus, the important conclusion we draw is that (14)
does not imply the vanishing of $all$ of the amplitudes
of $V_5$ but only of those not containing the operators
having singular OPE's with $\Gamma$.
For example, all  the correlators of the type
$<V_5(z_1)...V_5(z_m) Z_1(u_1)...Z_n(u_n)>$ 
where $Z_i(u_i)$ are perturbative superstring vertices (e.g. a photon)
must certainly vanish.
However, this does not imply the triviality of $V_5$.
Consider an operator 
\eqn\lowen{
W_5=H_{m_1...m_5}\oint{{dz}\over{2i\pi}}{e^\phi}
\psi^{m_1}...\psi^{m_5}e^{ikX}(z)}
 which is similar to the five-form
(5) with $e^{-3\phi}$ replaced by the operator $e^\phi$ of the identical
conformal dimension $-{3\over2}$.
Again, although this operator isn't 
manifestly BRST-invariant, below we shall demonstrate
how its invariance can be restored by adding the $b-c$ ghost dependent
terms. Its OPE with $\Gamma$ is given by
\eqn\lowen{\Gamma(z){e^\phi}
\psi^{m_1}...\psi^{m_5}e^{ikX}(w)\sim(z-w)^{-2}
e^{2\phi}\psi^{m_1}...\psi^{m_4}
(i(k\psi)\psi{m_5}+\partial{X_{m_5}})e^{ikX}(w)+...}
i.e. is precisely what we are looking for.
So our goal now is to derive the correction terms restoring the 
BRST-invariance  of $W_5$ and to demonstrate the nonzero correlator
involving the $V_5$ and $W_5$ operators. This would be a sufficient
proof of the non-triviality of both $V_5$ and $W_5$.
We start with the BRST invariance restoration.
The strategy is the following.
Consider the BRST charge given by
\eqn\lowen{Q_{brst}=\oint{{dz}\over{2i\pi}}(cT-bc\partial{c}
-{1\over{2}}\gamma\psi_m\partial{X^m}-{1\over4}\gamma^2b)}
Introduce an operator
\eqn\lowen{L(z)=-4ce^{2\chi-2\phi}\equiv:\xi\Gamma^{-1}:},
satisfying $\lbrace{Q_{brst}},L\rbrace=1$
Consider a non-invariant operator $V$
satisfying $\lbrace{Q_{brst},V}\rbrace=W$
for some $W$. Then, as $W$ is BRST-exact, clearly the
transformation $V{\rightarrow}V_{inv}=V-LW$ restores BRST-invariance,
provided that $V_{inv}$ is not exact (the latter 
has of course to be checked separately).
 Applying this scheme for $W_5$ of (17), we have
\eqn\grav{\eqalign{\lbrack{Q_{brst}},W_5\rbrack=
H_{m_1...m_5}(k)\int{{dz}\over{2i\pi}}
e^{2\phi-\chi+ikX}R_1^{m_1...m_5}(z)+be^{3\phi-2\phi+ikX}
R_2^{m_1...m_5}(z)
}}
where
\eqn\grav{\eqalign{R^{m_1...m_5}(z)=
-{1\over2}\psi^{m_1}...\psi^{m_5}(\psi\partial{X})
-{1\over2}\psi^{{\lbrack}m_1}..\psi^{m_4}(\partial^2{X^{m_5\rbrack}}
+\partial{X^{m_5\rbrack}}(\partial\phi-\partial\chi))
\cr-{i\over2}\psi^{m_1}...\psi^{m_5}
(k\psi)(\partial\phi-\partial\chi)+(k\partial\psi)}}
and
\eqn\lowen{R_2^{m_1...m_5}(z)=-{1\over4}(2\partial\phi-2\partial\chi-
\partial\sigma)\psi^{m_1}...\psi^{m_5}}
The next step is to cast this commutator as
\eqn\grav{\eqalign{\lbrack{Q_{brst},W_5\rbrack}\cr=
{1\over2}H_{m_1...m_5}(k)\oint_w{{dz}\over{2i\pi}}
(z-w)^2\partial^2_z(e^{2\phi-\chi}R_1^{m_1...m_5}+e^{3\phi-2\chi}
R_2^{m_1...m_5}(z))}}
where w is some worldsheet point; the expression (24) can obviously be
brought to the form (21) by partial integration. The contour integral
is taken around $w$; as the choice of $w$ is arbitrary, any
dependence on it in shall disappear in the end in correlation functions.
The next step is to insert the L-operator (20) inside the integral
(24). Evaluating the OPE of $L$ with the integrand of (24)
we obtain
\eqn\grav{\eqalign{
W_{5inv}(k,w)=H_{m_1...m_5}(k)\lbrace
\oint{{dz}\over{2i\pi}}e^{\phi}\psi^{m_1}...\psi^{m_5}e^{ikX}
\cr-{1\over2}\oint_w{{dz}\over{2i\pi}}(z-w)^2
:L\partial^2_z(e^{2\phi-\chi}R_1^{m_1...m_5}+e^{3\phi-2\chi}
R_2^{m_1...m_5}(z)):\rbrace
\cr=H_{m_1...m_5}(k)\lbrace
2\oint{{dz}\over{2i\pi}}e^{\phi}\psi^{m_1}...\psi^{m_5}e^{ikX}
-2\oint_w{{dz}\over{2i\pi}}(z-w)^2c{e^\chi}R_1^{m_1...m_5}(k,z)
\rbrace}}
This concludes the construction of the BRST-invariant 5-form state
at picture $+1$.
However one  still has to check if $W_5$-operator is non-trivial.
This is the legitimate concern since the BRST-invariant integrand
of (25) is BRST-invariant and has conformal dimension 3 and it is well-known
that invariant operators of conformal dimension other than 0 can 
always be expressed as BRST commutators.
Indeed, the commutator of any operator of dimension h with the zero mode of 
$T$ satisfies
\eqn\lowen{\lbrack{T_0},V_h\rbrack=hV_h}
where $T_0=\oint{{dz}\over{2i\pi}}zT(z)$.
Since $T_0=\lbrace{Q_{brst}},b_0\rbrace$, for any invariant $V_h$
one has 
\eqn\lowen{V_h={1\over{h}}\lbrace{Q_{brst}},b_0{V_h}\rbrace}
Acting on (25) with $b_0$ gives
\eqn\lowen{\eqalign{b_0W_{5inv}=
{2\over3}\oint_w{{dz}\over{2i\pi}}(z-w)^3{e^\chi}R_1^{m_1...m_5}(k,z)}}
accordingly $W_{5inv}$ can be represented as commutator
\eqn\grav{\eqalign{W_{5inv}=\lbrace{Q_{brst}},
{1\over6}\oint_w{{dz}\over{2i\pi}}(z-w)^{3}\xi{R_1^{m_1...m_5}}(k,z)
\rbrace}}
i.e. it is the BRST commutator with the operator $outside$ the
small Hilbert space. For this reason, the operator $W_{5inv}$
is BRST nontrivial.
The commutator (29) also ensures that the $W_{5inv}$ operator
is in the $small$ Hilbert space despite its
manifest dependence on $\xi$ since $\lbrace{Q_{brst}},\xi\rbrace$
is just the picture-changing operator.
An example of non-zero correlation function is a 3-point correlator
of 2 five-form states and one photon.
As it is clear from the above, one five-form should be taken 
at a picture +1 and another at a picture -3,
as any 3-point correlator containing the same-picture 
five-forms would vanish due to (14).
Technically, this means  that a photon state appears
in the  operator product
of two five-forms of positive and negative pictures,
but not in the OPE of the same picture five-forms.
The non-zero correlator involves one unintegrated
picture $+1$ photon: 
\eqn\lowen{
V_{ph}(q,z)=A_m(q):\Gamma{\lbrace}c(\partial{X^m}+i(k\psi)\psi^m)+\gamma\psi^m
{\rbrace}e^{iqX}:(z)}
 one unintegrated picture $-3$ five-form:
\eqn\lowen{V_5(k,z)=H_{m_1...m_5}(k)
:ce^{-3\phi}\psi^{m_1}...\psi^{m_5}e^{ikX}:(z)}
and one integrated five-form 
$W_{5inv}(p)$ of (25)
at picture $+1$.
Such a combination of pictures
 ensures the correct ghost number balance of the correlator
necessary to cancel the background ghost charges -
that is, the $\phi$ ghost number $-2$, $\chi$ ghostnumber $+1$ and
$b-c$ ghost number $-3$, due to the contribution of the correction term
$\sim{c\xi{R_{1}^{m_1...m_5}(k,z)}}$ of $W_5$.
Such a three-point function is in some way unusual
as the usual perturbative three-point correlators in string theory
normally involve the unintegrated vertices only.
Evaluating the correlator and evaluating the contour integral
around the insertion point $z_2$ of $W_{5inv}$ we find the result
to given by
The result of the computation is given by
\eqn\grav{\eqalign{
<V_5(k,z_1)W_{5inv}(p,z_2)V_{ph}(q,z_3)>
=H_{m_1...m_5}(k)H^{m_1...m_5}(p)((kq)(pA)-(pq)(kA))}}
After the Fourier transform,
this correlator leads to the low-energy effective action term given by
\eqn\lowen{S_{eff}\sim\int{d^{d}}X{(dH)_{l_1...l_5}^m}
{(dH)^{l_1...l_5{n}}}F_{mn}}
Remarkably, this term, originating from the worldsheet
correlator of two five-forms with a photon on a sphere
has the structure identical to the one obtained from the $disc$ amplitude
of a photon with
 two Ramond-Ramond operators of the $6$-form field strengths
$dH$. Thus the $V_5$ open string five-form state can be interpreted as a
source of the Ramond-Ramond $5$-form charge, i.e. a $D4$-brane.
This calculation also illustrates the 
 BRST non-triviality of $V_5$ and $W_5$ states.
Having illustrated the appearance of the picture-dependent
physical vertex operators, we are now prepared to give a
formal definition of ghost cohomologies.

1)The positive ghost  number  $N$ cohomology $H^{}_{N}(N>0)$ 
is the set of physical (BRST invariant and non-trivial)
vertex operators that 
 exist at positive superconformal ghost pictures
$n{\geq}N$ and that are annihilated by the inverse picture-changing operator
$\Gamma^{-1}=-4c\partial\xi{e^{-2\phi}}$ at the picture $N$.
 $:\Gamma^{-1}V^{(N)}:=0$
This means that the picture $N$ is the minimum positive picture
at which the operators $V\subset{H^{}_N}$ can exist.

2)The negative ghost number $-N$ cohomology $H^{}_{-N}$
consists of the physical vertex operators that exist at
negative superconformal pictures $n{\leq}{-N}(N>0)$ and that are annihilated 
by the direct picture changing at maximum negative picture $-N$:
$:\Gamma{V^{(-N)}}:=0$.

3)The operators existing at all pictures, including picture zero,
at which they decouple from superconformal ghosts,
are by definition the elements of the zero ghost cohomology
$H_0$.
The standard string perturbation theory thus involves
the elements of $H_0$, such as a photon.
 The picture $-3$ and picture $+1$ five-forms considered above 
are the elements of $H_{-3}^{}$ and $H^{}_{1}$ respectively.

4) Generically, there is an isomorphism between the positive and
the negative 
ghost cohomologies: $H_{-N-2}^{}\sim{H_{N}^{}};N{\geq}1$,
as the conformal dimensions of the operators
$:e^{-(N+2)\phi}(z)$ and $e^{N\phi}$ are equal and given by
$-{{N^2}\over2}-N$.
That is, to any element of $H_{-N-2}$ there corresponds an element
from $H_{N}$, obtained by replacing
$e^{-(N+2)\phi}\rightarrow{e^{N\phi}}$ and adding the 
appropriate $b-c$ ghost terms in order
to restore the BRST-invariance, using the L-operator (20).
For this reason, we shall refer to the cohomologies
 $H_{-N-2}^{}$ and ${H_{N}^{}}$ as dual.
With some effort, one can  also work out the precise isomorphism
relation between $H_{N}$ and $H_{-N-2}$
That is, let the integrated vertex operator $h_{-N-2}\equiv
\oint{{dz}\over{2i\pi}}W_{-N-2}(z)\subset
H_{-N-2}$ be the element of $H_{-N-2}$, with the integrand $W_{-N-2}(z)$
being some dimension 1 operator (for simplicity, we take it at 
tis maximal negative picture $-N-2$). The unintegrated version of this
operator is given by $cW_{-N-2}$
Then the corresponding operator, the element of $H_N$: 
$h_N\equiv\oint{{dz}\over{2i\pi}}W_N(z)\subset{H_N}$
can be constructed as
$$h_N=:\Gamma:^{2N+2}:ZcW_{-N-2}:$$
where 
$$Z(u)=\oint{{dw}\over{2i\pi}}(u-w)^3(bT+4c\xi\partial\xi{e^{-2\phi}}T^2)(w)$$
(with $T$ being the full matter$+$ghost stress-energy tensor of
the RNS theory)
is non-local, BRST-invariant and non-trivial $Z$-operator
mapping unintegrated vertices into integrated ones.
The properties of this operator have been discussed in ~{\self}
and basically, it can be viewed as a $b-c$ analogue
of $\Gamma$ (from this point of view,  integrated and unintegrated
vertices are just two versions of the same operator at different
$b-c$ pictures).
Since $\Gamma$ is invariant and non-trivial, 
$h_N$ is physical by construction,
once $h_{-N-2}$ is invariant and non-trivial.
The $L$-operator prescription for the five-form, described above,
is simply one explicit example of building such an isomorphism.
Technically, it is the non-local structure of $Z$
that leads to the terms from higher
Fourier modes in the expression (25) for $W_{5inv}$,
though such terms usually don't appear  in expressions
 for standard vertex operators, such as a photon.
Strictly speaking, the isomorphism between $H_{N}$ and $H_{-N-2}$ 
cannot be reduced to a usual
picture equivalence (though it is very much reminiscent of the one)
since the operators $Z$ and $\Gamma$ do not commute.
Note that $H^{}_{-1}$ and $H^{}_{-2}$ are empty, as any operator
existing at pictures $-2$ or $-1$ is either trivial or
can always be brought to picture zero
by $\Gamma$.

The important property of the ghost cohomologies is that
typically, the elements of $H_{-N}^{}$ have singular
operator products with the inverse picture-changing $\Gamma^{-1}$
while the elements of $H_{N}^{}$ have singular OPE's with
the direct picture-changing $\Gamma$.
As has been explained above, it is this property that ensures
that  correlators containing the vertices annihilated
by direct and inverse picture-changing, do not vanish despite (14),
provided that the correlator contains at least two vertices 
from the dual cohomologies of opposite signs. At least one of these vertices 
must be taken in the integrated form, for the reason 
we have pointed out above.
In critical strings, the only nontrivial cohomologies
are $H_{-3}^{}$ and its dual $H_{1}^{}$, i.e. the five-form states
(5), (25).
In non-critical strings, because of the Liouville dressing,
cohomologies of higher ghost numbers may appear as well.
In the next section we will show that the dimension 1 currents
from the higher ghost cohomologies enhance the symmetry algebra
of the target space and thus can be regarded as the generators
of hidden space-time symmetries originating from extra dimensions.

\centerline{\bf Ghost cohomologies, hidden symmetries and 2T-physics}

The properties of the picture-dependent five-forms (5),(25)
and their closed string counterparts have been studied
in a number of papers in the past (see e.g. ~{\selff}, ~{\selfian},
~{\sellf})
In particular, we have been able to show the relevance to the
non-perturbative dynamics of $D$-branes. For example,
the picture-dependent operators of the closed string sector
can be understood as the creadtion operators for $D$-branes
in the second quantized formalism ~{\selfian}.
This particularly means that the picture-dependent vertices
of the type (5), (25) (more precisely, their closed string counterparts)
describe the non-perturbative processes like an emission of a $D$-brane
by a string. Of course, once such a $D$-brane has been created
``ex nihilo'', this would lead to strong deformations of the space-time 
geometry. For instance, if an emission took place in the originally
flat space-time, the D-brane appearance
in the vacuum would result in strong fluctuations of
space-time metric which would gradually stabilize to the appropriate
D brane-type configuration. The processes of this type, describing
the evolution of the space-time metric from  flat to curved D-brane
 geometry, have been studied in the past by analyzing the worldsheet RG 
equations for sigma-models with the picture-dependent operators,
and it has been realised that generally these metric fluctuations
are described by either the stochastic equations of the Langevin type
with the non-Markovian noise or, in more complicated cases,
the Navier-Stokes type equations of hydrodynamics ~{\sellf}.
In this context, the flat and the curved D-brane metric 
can be viewed as different 
thermodynamical limits of the above stochastic processes.
Although it has been understood that the 
appearance of picture-dependent vertices strongly reshapes
the geometry of space-time,  
 certain general  principle, 
 relating the picture-dependent states to
 deformations of geometry (including the possible appearance
of new space-time symmetries and extra dimensions)
 has been lacking so far.
In this section we shall attempt to formulate such a principle
by studying the example of the ghost cohomologies
in non-critical NSR string theories with the Liouville dressing.
The particular result that we shall demonstrate below is the surprising
relation between the  ghost cohomologies  and the
extra symmetries from hidden space-time 
dimensions, observed in Bars 2T theories.
For example, one particular observation made by Bars is that
 the full symmetry group of a particle in  the $AdS_d$
space  is given by
$SO(d,2)$ which is larger than
the naive $SO(d-1,2)$ isometry group of the $AdS_d$ space
~{\bars}.
It has been shown by Bars that the generators of these
extra symmetries originate from hidden dimensions,
one of which is time-like ~{\bars}, ~{\barss}.
In this section, we shall discuss the string-theoretic
analogue of these new symmetries. 
Namely, we will show that the off-shell transformations in the 2T theories
observed by Bars et.al ~{\bars}, ~{\barss}, ~{\barsss},  
leading to non-linear symmetry transformations with their origin
in hidden space-time dimensions,
are produced by generators isomorphic
to the picture-dependent currents of higher ghost cohomologies.
Moreover, we shall argue that the off-shell symmetries 
and extra dimensions, observed by Bars ~{\bars}, ~{\barss}, ~{\barsss}
may not be complete as they are isomorphic to operators of the lowest nonzero 
ghost cohomologies; thus ``switching on'' the cohomologies of higher
ghost numbers may result in the extra symmetries not yet observed.
 The main idea behind the relation of ghost cohomologies to 
extra space-time symmetries is quite simple. In superstring theory,
the generators of space-time symmetries are given by the worldsheet integrals
of primary fields of dimension 1. For example, in flat space-time 
the dimension 1 operators $T^m=\oint{{dz}\over{2i\pi}}
{\partial{X^m}}, L^{mn}=\oint{{dz}\over{2i\pi}}\psi^m\psi^n;m=0,...d-1$
generate $d$ translations and ${1\over2}(d-1)d$ rotations of the 
Poincare group. These generators can of course be taken at 
equivalent ghost pictures; for instance at picture $-1$ we have
$T^m=\oint{{dz}\over{2i\pi}}e^{-\phi}\psi^m; L^{mn}=\oint{{dz}\over{2i\pi}}
c{\partial\xi}e^{-2\phi}\psi^{{m}}\psi^{n}$.
The current algebra of these operators generates the space-time symmetry
group  and the manifest space-time symmetries are therefore 
in one to one correspondence with
the dimension one primary operators of  ghost number  zero cohomology
with zero momenta. The appearance of the primary dimension 1 fields from
cohomologies of nonzero ghost numbers, however, 
extends the current algebra and 
thus leads to  new  transformations in space-time,
which are the candidates for new space-time symmetries
(the latter should be checked separately, upon the construction
of the generators). 
We shall demonstrate that these extra generators correspond to the hidden 
space-time symmetries, observed by Bars.
As an example, we will
consider the  noncritical superstrings in
$d-1$ dimensional Minkowski space-time ($d\geq{1}$).
It is easy to check that in this theory
the dimension 1 primary fields (including the matter and the Liouville
coordinates) with momentum zero in the $X$-direction
generate $SO(d-1,2)$ current algebra, which is 
 identical to the isometry of  $critical$ strings in $AdS_d$.
That is,
let $\varphi$ be the Liouville field and $\lambda$ its worldsheet 
superpartner.
Then it's easy to check that the dimension 1 primaries generating
$SO(d-1,2)$ are given by (in the limit of zero cosmological constant):
\eqn\grav{\eqalign{L^{mn}=\oint{{dz}\over{2i\pi}}\psi^{m}{\psi^n}\cr
L^{+m}=\oint{{dz}\over{2i\pi}}e^{-\phi}\psi^m\cr
L^{-m}=l(d)\oint{{dz}\over{2i\pi}}e^{Q\varphi}\psi^m\lambda\cr
L^{+-}=l(d)\oint{{dz}\over{2i\pi}}e^{-\phi+Q\varphi}\lambda}}
where $l(d)$ are the normalization constants for the Liouville-dependent 
operators, needed to compensate for the finite renormalization
of the structure constants due to the Liouville dressing ~{\dress}
(see the discussion below).
The stress-energy tensor and the supercurrent for the super Liouville system
are given by

\eqn\grav{\eqalign{T_{\varphi,\lambda}=-{1\over2}(\partial\varphi)^2
+{Q\over2}\partial^2\varphi-{1\over2}\partial\lambda\lambda\cr
G_{\varphi, \lambda}=-{1\over{{\sqrt{2}}}}
(\lambda\partial\varphi+Q\partial\lambda)}}

and the value of the background charge in $d$ dimensions
is given by

\eqn\lowen{Q={\sqrt{{{9-d}\over{2}}}}}

is chosen so that the total central charge of the system is zero:

\eqn\lowen{c_X+c_\psi+c_\varphi+c_\lambda+c_{b-c}+c_{\beta\gamma}=0}
and thus the conformal dimension of $e^{Q\varphi}$ is zero.
Next, $\eta^{mn}$ is the Minkowski metric, while
$\eta^{+-}=-1;\eta^{++}=\eta^{--}=0$.

The dressing with the dimension zero $e^{Q\varphi}$ operator
is necessary in order to ensure that the Liouville dependent currents are the 
primary  fields, due to the second derivative term in the Liouville stress
tensor.
Evaluating the commutators of the Liouville-dependent generators,
one has to compute their OPE's without the dressing first,
and then to dress the obtained operator on the right-hand side,
following the procedure similar to ~{\dress}.
As a result of the dressing, the structure constants are renormalized
by the factor of ${{k(d)+2}\over{k(d)+1}}$ ~{\dress}
where $k(d)$ is the central charge of $SL(2,R)$ current algebra 
in super Liouville theory
given by ~{\dress}:
\eqn\lowen{k(d)={1\over{8}}(d-1\pm{\sqrt{(d-1)(d-9)}})}
Accordingly, the normalization constants $l(d)$ of (34)
are to be chosen as
\eqn\lowen{l(d)={\sqrt{{k(d)+1}\over{k(d)+2}}}}
Our goal now is to demonstrate that for all values of $d$ (except for 
the special cases of the $c=1$ theory with $d-1=1$ 
and critical superstrings with $d-1=10$)
the picture-dependent dimension 1 primaries from $H_1\sim{H}_{-3}$
are given by $d+1$ generators, extending
the $SO(d-1,2)$ $AdS_d$ isometry to  $SO(d,2)$.
Such an enhancement of the space-time symmetry group
is precisely the one observed by Bars ~{\bars} for a particle
in $AdS_d$ due to  $d+1$ extra generators which, in the interpretation
of ~{\bars} originate from hidden space-time dimensions of the 2T-theory
living in $d+2$ dimensions, in which
 the $d$-dimensional universe is embedded.
As in ~{\bars}, these extra operators generate the nonlinear
transformations of the worldsheet Lagrangian
shifting it by a total derivative, i.e. leaving the action invariant.
In this context, the two extra dimensions observed by Bars admit 
a simple interpretation: one of them is the Liouville field,
another is the cohomology of the lowest nontrivial number.
As we have mentioned before, it is natural to assume that
the higher dimensional cohomologies lead to further  non-linear
space-time symmetries, so far  unobserved in the 2T physics formalism.
We start from reviewing the case of $d=2$ (supersymmetric $c=1$ model)
and then generalise it for higher values of $d$.
The case of $c=1$ model is special since, even without
the operators from nonzero ghost cohomologies, it contains
3 $SU(2)$ space-time generators, well-known to generate the
set of $SU(2)$ multiplet discrete states ~{\lian}, ~{\pol}, ~{\wit}:

\eqn\grav{\eqalign{T_{0,0}=\oint{{dz}\over{2i\pi}}
\partial{X}\cr
T_{0,1}=\oint{{dz}\over{2i\pi}}
e^{iX}\psi\cr
T_{0,-1}=\oint{{dz}\over{2i\pi}}e^{-iX}\psi}}

where the left index refers to the ghost number and the right
to the integer momentum in the $X$-direction.
It has been shown ~{\selfdisc} that the currents from 
$H_1\sim{H_{-3}}$ extend the $SU(2)$ current algebra
to $SU(3)$. Five extra generators from $H_1\sim{H_{-3}}$ are given by
~{\selfdisc}:
\eqn\grav{\eqalign{T_{-3,2}=\oint{{dz}\over{2i\pi}}e^{-3\phi+2iX}\psi(z)\cr
T_{-3,1}=\oint{{dz}\over{2i\pi}}e^{-3\phi+iX}(\partial\psi\psi
+{1\over2}(\partial{X})^2+{i\over2}\partial^2{X})(z)\cr
T_{-3,-1}=\oint{{dz}\over{2i\pi}}e^{-3\phi-iX}(\partial\psi\psi
+{1\over2}(\partial{X})^2-{i\over2}\partial^2{X})(z)\cr
T_{-3,0}=\oint{{dz}\over{2i\pi}}e^{-3\phi}(\partial^2{X}\psi-2\partial{X}
\partial{\psi})(z)\cr
T_{-3,-2}=\oint{{dz}\over{2i\pi}}e^{-3\phi-2iX}\psi(z)
}}
It is straightforward to check that these operators are
the BRST-invariant and non-trivial primary fields of dimension
one, annihilated by $\Gamma$ and  are thus the elements of $H_{-3}$.
Of course, the $H_{1}$-version of these operators can be constructed as well
by replacing $e^{-3\phi}$ with $e^\phi$ and adding the $b-c$ ghost terms
by using the L-operator prescription (20).

The $SU(3)$ algebra is 
then generated by combining (41) with 3 operators (40) of $SU(2)$
The Cartan subalgebra of $SU(3)$ is formed by
the zero momentum generators
$T_{0,0}$ and $T_{-3,0}$ while 3 operators with negative and 3 with
 positive momenta form the lower and upper nilpotent subalgebras respectively.
The SU(3) multiplet states are then obtained by acting
with various combinations of the lowering operators
$T_{-3,-1},T_{-3,-2}$ and $T_{0,-1}$
on the dressed tachyonic highest weight vectors with positive integer momenta.
The  prescription to obtain these operators is the following:
one starts with maximal positive momentum generator $T_{-3,2}$ 
of $H_{-3}$ and
acts on it repeatedly with the lowering $T_{0,-1}$ of $SU(2)$,
 descending down to the maximal negative momentum generator
$T_{-3,-2}$ (which in turn is annihilated by $T_{0,-1}$) thus generating
5 currents $T_{-3,n}$ of $H_{-3}$ $(-2\leq{n}\leq{2})$.
The described procedure is straightforward to generalize  for any
higher ghost number hohomology $H_{-n}\sim{H_{n-2}};n\geq{3}$.
That is, one starts with the maximum positive momentum $H_{-n}$ primary
$T_{-n,n-1}=\oint{{dz}\over{2i\pi}}e^{-n\phi+i(n-1)X}$
and then generates $n+1$ $T_{-n,m}$ generators of $H_{-n}$
or its dual $H_{n-2}$  $(-n\leq{m}\leq{n})$
by repeated action on $T_{-n,n-1}$ with $T_{0,-1}$.
Then, combining together all the operators from
$H_{-i}\sim{H_{i-2}};i=0,3,4,...,n$ one obtains
$n^{2}-1$ currents generating $SU(n)$, although the explicit
expressions for the generators become quite cumbersome for $n>3$
~{\selfdisc}
. Again, the $SU(n)$ multiplet states are obtained 
by acting
on the dressed tachyonic primaries
with integer positive momenta with various combinations of $T_{-n,m}
\sim{T_{n-2,m}}$ with $m<0$. Their structure constants
are given by the Clebsch-Gordan coefficients of $SU(n)\sim{SL}(n,R)$ (up to 
the factors - functions of the Casimir eigenvalues) 
and are thus related to the volume 
preserving diffeomorphisms in $N$ dimensions $SDiff(n)$.
Remarkably, this implies that each time we ``switch on'' 
the  generators from a cohomology of the higher ghost number,
we ``open up'' an extra space-time dimension.

Let us now consider the case of non-critical strings in higher dimensions.
The direct generalization of the procedure, described for
the case of $c=1$, doesn't work  for $d-1>1$,
as we no longer have generators with the discrete
nonzero momenta. For example, the analogue of the current
$T_{-n,n-1}$ is now the operator $\oint{{dz}\over{2i\pi}}e^{-n\phi+ik_m{X^m}}$
where the momentum ${\vec{k}}$ satifies $k_mk^m=n-1$, 
but is arbitrary otherwise.
For this reason,   the only $H_{-n}\sim{H}_{n-2}$-operators $(n\neq{0,1,2})$
that the higher dimensional strings could ``inherit'' from the $c=1$ case
are those carrying momentum zero. That is, they should inherit the structure
of the Cartan generators of $SU(n)$.
Consider the case $n=3$.
The  straightforward
analogue of the $T_{-3,0}$ Cartan generator of $SU(3)$
is the trace operator
\eqn\lowen{
L^{+\alpha}=\oint{{dz}\over{2i\pi}}e^{-3\phi}(\partial^2{X_m}\psi^m-
2\partial{X_m}\partial\psi^m)}
where by definition $\alpha=1,\eta^{\alpha\alpha}=1$.

The Greek space index $\alpha$ corresponds to what will turn out to be an 
extra dimension, induced by the generators of $H_{-3}\sim{H_1}$;
of course, it should be distinguished from $m=1$ of the $(d-1)$-dimensional
Minkowski space.
Note that the generators 
\eqn\lowen{\rho_m^{n}=\oint{{dz}\over{2i\pi}}e^{-3\phi}(\partial^2{X_m}\psi^n-
2\partial{X_m}\partial\psi^n)}
 are $not$ the elements
of $H_{-3}$ for $n\neq{m}$ since they are BRST-exact (except for $n=m$).
That is, it is straightforward to check
that
\eqn\lowen{\rho_m^n=\lbrack{Q_{brst}},:\Gamma^{-1}U_m^n:\rbrack}
with 
\eqn\lowen{
U_m^n=\oint{{dz}\over{2i\pi}}\partial{\xi}e^{-3\phi}\psi_m\partial{X^n}}
On the other hand,the generator $L^{+\alpha}\equiv{Tr}{\rho_m^n}$
is nontrivial as for the coincident $m$ and $n$
$\lbrack{Q_{brst},U_{m}^m}\rbrack=0$.
Next, using the Liouville field $\varphi$ and its worldsheet superpartner
$\lambda$ it is not difficult to construct other $d$ $H_{-3}$-generators
of the $T_{-3,0}$-type.
That is, it is straightforward to check that
$d-1$ generators

\eqn\grav{\eqalign{L^{m\alpha}=l(d)\oint{{dz}\over{2i\pi}}
e^{-3\phi+Q\varphi}\lbrace(\partial^2\varphi+Q(\partial\varphi)^2)\psi^m
-2\partial\varphi\partial\psi^m+\partial^2{X^m}\lambda\cr
-2\partial{X^m}
\partial\lambda-4Q\partial\varphi\lambda\partial{X^m}\rbrace}}
are the BRST-invariant and non-trivial primary fields of dimension 1,
annihilated by $\Gamma$, i.e. are the elements of $H_{-3}\sim{H_1}$
The final generator of the $T_{-3,0}$-type, the element of $H_{-3}\sim{H_1}$
 can be constructed using just
the superconformal ghosts, $\varphi$ and $\lambda$.
It is given by 
\eqn\lowen{L^{-\alpha}=l(d)\oint{{dz}\over{2i\pi}}e^{\phi+Q\varphi}
\lbrace(\partial^2\varphi+Q(\partial\varphi)^2)\lambda+
\partial\varphi\partial\lambda\rbrace}
The final step is to check that $d+1$ generators
$L^{+\alpha},L^{-\alpha}$ and $L^{m\alpha}$ of $H_{-3}$, along with
${1\over2}(d-1)d$ generators of $SO(d-1,2)$ combine into
${1\over2}d(d+1)$ generators of $SO(d,2)$

Introducing the $d+2$-dimensional index
$M=(m,+,-,\alpha);m=0,...{d-2};\alpha=1$ with the $(d,2)$ metric
$\eta^{MN}$ consisting of $\eta^{mn},\eta^{+-}=-1,
\eta^{--}=\eta{++}=0,\eta^{\alpha\alpha}=1$
and evaluating the commutators one can show that
\eqn\lowen{\lbrack{L^{M_1N_1},L^{M_2N_2}\rbrack}
=\eta^{M_1M_2}L^{N_1N_2}+\eta^{N_1N_2}L^{M_1M_2}
-\eta^{M_1N_2}L^{M_2N_1}-\eta^{M_2N_1}L^{M_1N_2}}
 As before, the commutators of the operators (46), (47)
must first be computed in the limit of zero dilaton field
(i.e. with the generators taken without the dressing)
with the subsequent Liouville dressing of the r.h.s. 
of the obtained commutator.
In the commutators (48) with the both of $L^{MN}$'s coming from
$H_{-3}\sim{H_1}$ one has to take  one of the generators
at picture $-3$ and another at picture $+1$, since only
the OPE's of the currents from the dual cohomologies
of the opposite sign ($H_1$ and $H_{-3}$)
contain the non-trivial operators (typically, those of $H_0$
). On the contrary,
the OPE of any 2 operators which are both from $H_{-3}$
or both from $H_{1}$ cannot contain any non-trivial
currents of $H_0$, as has been explained above.
 Structurally,
the commutator of two currents from $H_{-3}$ and $H_{1}$
gives the current from $H_0$ at picture $-2$.
The commutators of currents from $H_{-3}$ (or $H_1$) with
operators from $H_0$ typically produce the currents
of $H_{-3}$ (or $H_1$).
The $L^{MN}$-commutators constructed above are thus in one to one
 correspondence with the  L-generators of $SO(d,2)$ , observed
by Bars for the $AdS_d$ particle. The generators
from the nonzero ghost cohomologies  are in one to one correspondence with
the generators from hidden extra dimensions in Bars approach.

\centerline{\bf 2T-physics and $\alpha$-symmetry}

To conclude this paper, we will demonstrate that the generators
 $L^{-\alpha},L^{+\alpha}$ and $L^{m\alpha}$ of $L^{MN}$ generate 
the nonlinear
transformations of the worldsheet $NSR$ Lagrangian that
leave it invariant, up to total derivative.
These extra symmetries are  similar to the non-linear symmetry
transformations for $AdS_d$ particle, first observed by Bars.
We start with the symmetry transformations generated by
the trace operator $L^{+\alpha}$.
Applying the generator 
$\epsilon^{+\alpha}L^{+\alpha}$
to various $RNS$ fields (where $\epsilon^{+\alpha}$ is the 
infinitezimal parameter)
one easily finds
the corresponding transformations to be given by

\eqn\grav{\eqalign{\delta{X^m}=\epsilon^{+\alpha}
\lbrace\partial(e^\phi\psi^m)+2e^\phi\partial\psi^m\rbrace\cr
\delta\psi^m=\epsilon^{+\alpha}\lbrace{-}e^\phi\partial^2{X^m}
-2\partial(e^\phi\partial{X^m})\rbrace\cr
\delta\gamma=\epsilon^{+\alpha}{e^{2\phi-\chi}}(\psi_m\partial^2{X^m}
-2\partial\psi_m\partial{X^m})\cr
\delta\beta=\delta{b}=\delta{c}=0}}
and similarly for  the anti-holomorphic fields.

The next step is to show that the RNS worldsheet action:

\eqn\grav{\eqalign{S_{NSR}={1\over{2\pi}}\int{d^2z}
\lbrace{-}{1\over2}\partial{X_m}\bar\partial{X^m}-
{1\over2}\psi^m\bar\partial\psi^m-{1\over2}\bar\psi_m\partial\bar\psi^m\cr
+b\bar\partial{c}+{\bar{b}}\partial{\bar{c}}+\beta\bar\partial\gamma
+{\bar{\beta}}\partial{\bar{\gamma}}\rbrace+S_{Liouville}\cr
S_{Liouville}={1\over{4\pi}}\int{d^2z}
{\lbrace}\partial\varphi\bar\partial\varphi+\lambda\bar\partial\lambda+
{\bar\lambda}\partial\bar\lambda-F^2+2\mu_0be^{b\varphi}
(ib\lambda\bar\lambda-F)\rbrace}}
with
$$Q=b+{1\over{b}}$$
is invariant under the transformations (49).
Integrating by parts, we have:

\eqn\grav{\eqalign{\delta(-{1\over{4\pi}}\int{d^2z}
\partial{X_m\bar\partial{X^m}})={1\over{2\pi}}
\int{d^2z}\delta{X_m}\partial\bar\partial{X^m}\cr
={1\over{2\pi}}\int{d^2z}\lbrace\partial(e^\phi\psi_m)
+2e^\phi\partial\psi_m\rbrace\partial\bar\partial{X^m}
\cr
={1\over{2\pi}}\int{d^2z}\lbrace{-}e^\phi\psi_m\partial^2\bar\partial{X^m}
+2e^\phi\partial\psi_m\partial\bar\partial{X^m}\rbrace}}
Next, the transformation of the holomorphic 
$\psi_m\bar\partial\psi^m$-term of the worldsheet action (50)
under (49) is given by:
\eqn\grav{\eqalign{\delta(-{1\over{4\pi}}\int{d^2z}
\psi_m\bar\partial\psi^m)={1\over{2\pi}}
\int{d^2z}\bar\partial(\delta\psi_m)\psi^m
\cr
=-{1\over{2\pi}}\int{d^2z}\bar\partial
{\lbrace}e^\phi\partial^2{X_m}+2\partial(e^\phi\partial{X_m})\rbrace
\psi^m\cr
=-{1\over{2\pi}}\int{d^2z}\lbrace{\bar\partial}(e^\phi\partial^2{X_m})\psi^m
-2\bar\partial(e^\phi\partial{X_m})\partial{\psi^m}\rbrace}}
Adding these contributions together, we get
\eqn\grav{\eqalign{\delta{S_X}+\delta{S_\psi}\equiv\delta(-{1\over{4\pi}}
\int{d^2z}\lbrace\partial{X_m}\bar\partial{X^m}
+\psi_{m}\bar\partial\psi^m\rbrace)\cr
=-{1\over{2\pi}}\int{d^2z}(\bar\partial{e^\phi})\lbrace
\partial^2{X_m}\psi^m-2\partial{X_m}\partial\psi^m\rbrace}}
Finally, using the OPE relation
\eqn\lowen{lim_{z\rightarrow{w}}
:\beta:(z):{e^{2\phi-\chi}}:(w)\equiv
lim_{z\rightarrow{w}}:e^{\chi-\phi}\partial\chi:(z):e^{2\phi-\chi}(w):
=-:e^{\phi}:(z)}
and the fact that $\delta\beta=0$
we find the transformation of the $\beta-\gamma$ term of $S_{RNS}$
to be given by
\eqn\grav{\eqalign{\delta{S_{\beta\gamma}}\equiv
\delta({1\over{2\pi}}\int{d^2z}\beta\bar\partial\gamma)
=-{1\over{2\pi}}\int{d^2z}\bar\partial\beta\bar(\delta\gamma)\cr
={1\over{2\pi}}\int{d^2z}(\bar\partial{e^\phi})
\lbrace
\partial^2{X_m}\psi^m-2\partial{X_m}\partial\psi^m\rbrace,}}
i.e. 
\eqn\lowen{\delta({S_X}+S_\psi+S_{\beta\gamma})=0}
Since the $b-c$ and the Liouville parts of the RNS action (50)
are manifestly invariant under (49),
this means that $\delta{S_{RNS}}=0$ and the transformations
(49) constitute the new space-time symmetry
non-critical $RNS$ superstring theories.
We shall refer to this  new nonlinear symmetry of $S_{RNS}$
as the $\alpha$-symmetry and the corresponding
transformations (49) as $\alpha$-transformations, in order to emphasize
their relation to nonzero ghost cohomology generator $L^{+\alpha}\subset
{H_{-3}}\sim{H_1}$ of $SO(d,2)$ and, accordingly, to
the hidden space-time dimension, which
we have parametrized with the index $\alpha$ (see above).
In case of the critical ten-dimensional superstring theory 
 this is the only new additional symmetry of the RNS action
(apart from the usual translations and rotations).
For  non-critical RNS strings in $d-1$ dimensions,
 however, there are also $d$ additional $\alpha$-transformations,
generated by  the Liouville-dependent
$L^{-\alpha}$ and $L^{m\alpha}$ of $SO(d,2)$
($m=0,...,d-2$).
Proceeding exactly as above,
one can easily derive
 $d-1$ $\alpha$-symmetry transformations for $S_{RNS}$,
generated by $\epsilon_{m\alpha}L^{m\alpha}$, to be given by,
in the zero cosmological constant limit and in the absense of dilaton:
\eqn\grav{\eqalign{\delta{X_m}=\epsilon_{m\alpha}{\lbrace}
\partial(e^\phi\lambda)+2e^{\phi}\partial\lambda
\rbrace\cr
\delta\lambda
=-\epsilon_{m\alpha}{\lbrace}
2\partial(e^\phi\partial{X^m})+e^\phi
\partial^2{X^m}\rbrace\cr
\delta\gamma=\epsilon_{m\alpha}e^{2\phi-\chi}\lbrace
\partial^2{X^m}\lambda
-2\partial{X^m}\partial\lambda\rbrace\cr
\delta\beta=\delta{b}=\delta{c}=\delta\varphi=\delta\psi^m=0}}
The final, (d+1)-st  $\alpha$-symmetry transformation of $S_{NSR}$,
generated by $L^{-\alpha}$ of (47)  acts on the superconformal
ghosts and the Liouville field only and is given by

\eqn\grav{\eqalign{\delta\varphi
=\epsilon_{-\alpha}\lbrace
\partial(e^\phi\lambda)+2e^\phi\partial\lambda
\rbrace\cr
\delta\lambda=-\epsilon_{-\alpha}\lbrace{2}\partial(e^\phi\partial\varphi)
+e^\phi{\partial^2}\varphi\rbrace\cr
\delta\gamma=\epsilon_{-\alpha}e^{2\phi-\chi}\lbrace
\lambda\partial^2\varphi-2\partial\varphi\partial\lambda
\rbrace\cr
\delta\beta=\delta{b}=\delta{c}=\delta{X^m}=\delta\psi^m=0}}
Repeating the derivation we have performed above for
$L^{+\alpha}$, it is now easy to check that
the $\alpha$-transformations (57), (58) are the symmetries of the 
noncritical RNS. Thus the full $SO(d,2)$ symmetry group of
non-critical strings in $d-1$ consists of ${1\over{2}}d(d-1)$
transformations of $AdS_d$ isometry plus $d+1$ $\alpha$-transformations
generated by currents from nonzero ghost cohomology $H_{-3}\sim{H_1}$.
This non-linear $\alpha$-symmetry is thus the new matter-ghost mixing symmetry
of the RNS strings and must be the stringy analogue of the 
non-linear symmetries for the $AdS_d$ particle  from hidden
extra dimensions, previously observed in the $2T$-physics approach.

\centerline{\bf Discussion}

In this paper we have demonstrated that the appearance of the
new  physical ghost-dependent generators from
the first non-trivial ghost cohomology $H_{-3}\sim{H_1}$
in non-critical RNS superstring theories leads to the enhancement
of the current algebra of space-time generators
from $SO(d-1,2)$ to $SO(d,2)$
 and to new
non-trivial global symmetries in space-time, associated with
the new ghost-dependent generators. The  non-linear
symmetry transformations, called $\alpha$-transformations
in this paper, are the symmetries of the $RNS$ Lagrangian.
The generators of the $\alpha$-symmetry appear to be in one to one 
correspondence with the extra symmetry generators of the $2T$ theories
which, in the Bars approach, are related to the hidden
symmetries originating from higher space-time dimensions.
So in this sense, our results are the stringy extention
of the work by Bars. The $\alpha$-symmetry of the worldsheet
superstring action, discussed in this paper, has not been
previously observed and this by itself appears to be an important
finding.
The natural question is what happens if one considers the ghost
cohomologies of higher ghost numbers, for example,
$H_{2}{\sim}H_{-4}$. In the special, and the simplest case
of supersymmetric $c=1$ theory, each new cohomology
effectively corresponds to opening up a
theory to a  new hidden space-time dimension.
From the technical point of view,  the $c>1$ case
appears to be more complicated, but it is tempting to assume
that qualitatively the same scenario is still true for $c>1$:
that is, the extra space-time generators from $H_{n}\sim{H}_{-n-2};
n=2,3,...$ correspond to yet unknown 
space-time symmetries of the theory, related to yet unknown hidden
space-time dimensions.
Technically, the space-time symmetry generators
from $H_{n}\sim{H_{-n-2}}$ should inherit their structure from
the Cartan generators of $SU(n+2)$ of the $c=1$ case
- precisely as we have shown it for the $n=1$ case.
Unfortunately, for $n\geq{2}$ the expressions for the 
Cartan generators become 
quite cumbersome and we leave the $n\geq{2}$ case for the future work.
Nevertheless, it seems reasonable to believe that
the $\alpha$-symmetry transformations, derived in this paper,
are not at all the end of the story. Given the
correspondence between the  $H_1\sim{H_{-3}}$ 
$\alpha$-symmetry generators, and the 
additional  symmetries of $2T$-physics
in the case of a particle, the interesting question
is whether the list of new symmetries (and, accordingly,
the number of the hidden dimensions) observed by Bars, is complete.
Superstring theory seems to predict that this list is incomplete,
and new hidden symmetries, as well as new space-time dimensions,
are yet to be observed.

\listrefs
\end